\documentclass[doublecol]{epl2} 
\usepackage{nicefrac}
\usepackage{umlaut}
\usepackage{amsmath}

\title{Extending the scope of microscopic solvability: Combination of
  the Kruskal-Segur method with Zauderer decomposition}
\shorttitle{Combining Kruskal-Segur with Zauderer}

\author{Thomas Fischaleck \and  Klaus Kassner}
\shortauthor{Thomas Fischaleck \etal}
\institute{Institut f\"ur Theoretische Physik,
  Otto-von-Guericke-Universit\"at Magdeburg,
  \\
  Postfach 4120, D-39016 Magdeburg, Germany}

\date{October 24, 2007}


\abstract{%
  Successful applications of the Kruskal-Segur approach to interfacial
  pattern formation have remained limited due to the necessity of an
  integral formulation of the problem. This excludes nonlinear bulk
  equations, rendering convection intractable.  Combining the method
  with Zauderer's asymptotic decomposition scheme, we are able to
  strongly extend its scope of applicability and solve selection
  problems based on free boundary formulations in terms of partial
  differential equations alone.  To demonstrate the technique, we give
  the first analytic solution of the problem of velocity selection for
  dendritic growth in a forced potential flow.
}

\begin{document}


\maketitle

\newcommand{\Vau}{M}
\newcommand{\We}{N}

The fundamental equations describing the growth of a crystal into its
undercooled melt 
are very difficult to solve, if surface tension effects are accounted
for, even when restriced to the simplest case of merely diffusive heat
transport.  On the other hand, the capillary length $d_0$ describing
these effects is typically very small in comparison with other length
scales of the problem such as the sizes of growing patterns or the
diffusion length.  Therefore, it was a natural step to first look for
solutions with $d_0$ set equal to zero.  This simplified problem was
solved exactly by Ivantsov \cite{ivantsov47} who showed that the
crystal can grow in the shape of a parabola in 2D or a paraboloid in
3D.  A major drawback of these solutions is that they constitute a
whole continuum for any given undercooling: the mathematics fixes only
the Péclet number $P_c= V\rho/D$, where $V$ is the growth velocity of
the crystal, $\rho$ the tip radius of the parabolic needle, and $D$
the thermal diffusion coefficient.  Hence, only the product of
velocity and length scale is determined, but neither of the two
quantities separately. In experiments, a given undercooling leads to
both a well-defined growth velocity and a well-defined tip radius of
the needle crystal, which after developing side branches is called a
dendrite.  This situation became known as the {\em selection problem}
of diffusion-limited dendritic growth and is was not solved
until some twenty years ago 
\cite{caroli86,benamar86,barbieri87,brener91a}, 
with the advent of {\em microscopic solvability theory}.

Because the theory was mathematically complex and not very intuitive,
it failed to enjoy unanimous appraisal.
Moreover, its success in explaining experiments remained controversial
to some extent \cite{glicksman93}. It has been emphasized by Tanveer
\cite{tanveer00} that even small fluid flows in the melt might account
for changes in the theoretically predicted scalings as the problem is
structurally unstable.  Hence, selection theory should be extended to
nondiffusive transport such as convection.  To our knowledge, the only
approach to solvability theory available so far for models with
convection is due to Bouissou and Pelcé (BP) \cite{bouissou89}.  Their
method relies on a linearized solvability condition, which prevents it
from becoming exact in the limit of vanishing $d_0$.  Also, it has
been shown \cite{tanveer90} that nonlinearity may be crucial in
problems involving multiple parameters.
Hence, a method would be more than desirable that takes nonlinear
solvability into account but can also deal with problems  not
permitting an integral formulation. This letter 
presents such an approach.

Lengths are nondimensionalized by the tip radius $\rho$ of the
Ivantsov parabola \cite{ivantsov47}, which is then given by
$y-y_0=\frac{1}{2}-\frac{1}{2}x^2$ in a comoving coordinate system.
We use conformal parabolic coordinates $x=\xi\eta$,
$y-y_0=\frac{1}{2}(\eta^2-\xi^2)$, so the Ivantsov parabola reads
$\eta=1$.  The relevant nondimensional parameters of the problem are
the growth Péclet number $P_c$, defined above, the stability parameter
$\sigma$, and the flow Péclet number $P_f$.  The latter are given in
terms of physical quantities by $\sigma=2 d_0 D/[V\rho^2]$,
$P_f=U\rho/D$, where $U$ is the velocity of an imposed flow.

In order to demonstrate the power of the method, we first show how it
simplifies a 
solved problem, the growth of of a needle
crystal under purely diffusive heat transport.  Next we deal with
the hitherto unsolved selection problem of a crystal growing in a
potential flow, where the basic field equations are nonlinear.
We simplify the presentation using some
approximations that can in principle be avoided and will be removed in
a more extended version of this article.
The first problem was treated via the Kruskal-Segur method \cite{KS}
by Ben Amar et al. \cite{benamar86}.  They start from an
integral equation describing steady state growth in the limit $P_c\to
0$.  The integral term is linearized about the Ivantsov parabola,
setting $\eta_s=1+h$ [$\eta_s(\xi)$ is the interface position].
After reducing the integral to a local expression using sophisticated
complex analysis \cite{brener91a}, one finds the dominant
behavior of the solution near a singularity in the complex plane at
$\xi=-{i}$:  
\begin{equation}
\label{basic-}
\sigma\kappa=(1-{i}\xi)h(\xi)\>,
\end{equation}
where 
\begin{align}
\kappa=\frac{-1}{(\xi^2+(1+h)^2)^{\frac12}}
&\left\{\frac{h^{\prime\prime}}{(1+h^{\prime 2})^{\frac{3}{2}}}\right.\nonumber\\
 &\left.\mbox{} +\frac{\xi h^\prime-1-h}{(\xi^2+(1+h)^2) (1+h^{\prime
      2})^{\frac12}}\right\}
\end{align}
is the curvature [the prime denotes a
derivative w.r.t.~the argument (i.e.~$\xi$)]. Equation
(\ref{basic-}), a second-order nonlinear differential equation for
$h(\xi)$, contains all the information needed to compute the
transcendental corrections (i.e., the mismatch function) that have to
be suppressed at the needle tip for selection to be possible.
Essential for its derivation was the use of an integral
equation, available only for linear bulk equations.

We now rederive Eq.~(\ref{basic-}) from the differential
equation formulation of the free-boundary problem directly.  The field
equation for the problem linearized about the Ivantsov solution is
just the Laplace equation $\partial^2_\xi T+\partial^2_\eta T = 0$, 
both in the liquid and solid phases.  The interface
boundary conditions become:
\begin{eqnarray}
T|_s\mbox{} =\mbox{}-\frac{\sigma}{2}\kappa \>, \quad
T|_l&=&T|_s+h \>,\nonumber \\
h+\xi h^\prime+\left(\partial_\eta-h^\prime \partial_\xi\right)
\left(T|_l-T|_s\right)&=&0  \>.\label{balance}
\end{eqnarray}
Of these, the first equation is the Gibbs-Thomson relation, the second
expresses continuity of the temperature at the interface (subscripts
$l$ and $s$ refer to evaluation at the position of the Ivantsov
parabola), the third is the continuity equation.  Noting that
$(\partial_\xi^2+\partial_\eta^2)=
(\partial_\xi+{i}\partial_\eta)(\partial_\xi-{i}\partial_\eta)$,
we replace the bulk equation with
\begin{eqnarray}
(\partial_\xi+{i}\partial_\eta)T&=&0 \quad\mbox{in the liquid}\>,
\label{bulkl2}\\
(\partial_\xi-{i}\partial_\eta)T&=&0  \quad\mbox{in the solid}\>.
\label{bulks2}
\end{eqnarray}
From Eqs.~(\ref{balance}), (\ref{bulkl2}) and
(\ref{bulks2}) we get, neglecting terms quadratic in $h^\prime$ \cite{expla}
\begin{equation}
\Bigl[(1-{i}\xi) h\Bigr]^\prime
=-2\left(T|_s\right)^\prime=\sigma\kappa^\prime\>,\label{localeqn2}
\end{equation}
and after one integration, obtaining the integration constant from the
boundary condition $h\to 0$ $(\xi\to\infty)$, we recover
(\ref{basic-}) almost effortless.

Let us briefly discuss the philosophy of this approach.  The
temperature field satisfies the Laplace equation, solved by
$T(\xi,\eta)=f_1[\xi+i(\eta-1)]+f_2[\xi-i(\eta-1)]$. After inserting
this general solution into the boundary conditions (\ref{balance}),
valid at $\eta=1$, we analytically continue these to the vicinity of
$w=-i$ ($w$ is the analytic continuation of $\xi$).  Some of the
terms must become singular there to compensate for
the singularity of the curvature term. Since the solution for the
liquid must be analytic in the upper half plane ($\eta>1$), the $f_2$
term remains regular near $w=-i$ and is hence negligible. The other
term is a solution of (\ref{bulkl2}). Similarly, dropping regular
terms from the solid-side solution, we keep the term that solves
(\ref{bulks2}).  This procedure gives a valid approximation near the
singularity.  Far away from singularities it is also justified,
because the curvature term in Eqs.~(\ref{balance}) can be linearized
and the corresponding inhomogeneity ignored.
We also have to take care of the singularity at $w=i$. Since the
final result has to be real, this singularity leads to the complex
conjugate.

Let us determine the transcendental mismatch for later reference
when considering the case with flow. 
Far from the singularity, 
the homogeneous part of the linearized Eq.~(\ref{basic-}) reads
\begin{equation}
\sigma\left(h^{\prime\prime}+\frac{\xi}{1+\xi^2}
h^\prime\right)+\sqrt{1+\xi^2}(1-{i}\xi)h=0 \>. \label{WKBbasic-}
\end{equation}
This may be solved using WKB techniques, yielding an \emph{outer} WKB solution: 
\begin{eqnarray}
h&=&B (1-{i}\xi)^{-\frac{5}{8}}
(1+{i}\xi)^{-\frac{3}{8}}
\exp\left\{\sigma^{-\frac{1}{2}}S(\xi)\right\},\label{transcorr}
\\[-0.1cm]
S(\xi)&=&{i}\int\limits_{-{i}}^{\xi}
(1-{i}\xi^\prime)^{\frac{3}{4}}(1+{i}\xi^\prime)^{\frac{1}{4}}
\,\mathrm{d}\xi^\prime \>.\label{phaseWKB}
\end{eqnarray}
To obtain the \emph{inner} equation near $\xi=-{i}$, we rescale
\begin{equation}\label{rescaling}
h=\sigma^\alpha\phi \>, \quad 
\xi=-{i}+{i}\,\sigma^{\alpha} t \>.
\end{equation}
From equation (\ref{basic-}), we have
\begin{equation}
\sigma^{2\alpha}t\phi=\sigma^{1-\frac{3}{2}\alpha}K+
O\left(\sigma^{-\frac{1}{2}\alpha+1}\right) \,, \label{localeqnlower}
\end{equation}
where
\begin{equation}
K=\frac{1}{(2t+2\phi)^{\frac{1}{2}}}
\left\{\frac{\ddot{\phi}}{(1-\dot{\phi}^{2})^{\frac{3}{2}}}
+\frac{\dot{\phi}+1}{(2t+2\phi)(1-\dot{\phi}^{2})^{\frac{1}{2}}}\right\}.
\label{Kdef}
\end{equation}
To balance both sides of equation (\ref{localeqnlower}), we need
 $\alpha = \frac{2}{7}$, hence $t\phi=K$, and  
the asymptotic behavior of $\phi$ for
large $t$ is $\phi\sim 2^{-\frac{3}{2}}t^{-\frac{5}{2}}$.
Linearizing about the asymptotic solution and performing a local
asymptotic analysis near $t=\infty$, we get the decreasing
eigenfunction
$g=t^{-\frac{5}{8}}
\exp\left\{-\frac{4}{7}2^{\frac{1}{4}}t^{\frac{7}{4}}\right\}$,
leading to the transcendental behavior
\begin{equation}
\phi = 2^{-\frac{3}{2}}t^{-\frac{5}{2}}+A\,  
g(t)\,\,\,\,\,\,\,(t\rightarrow\infty)\>,\label{Aeqn}
\end{equation}
where the nonlinear eigenvalue $A$ can be found by solving $t\phi=K$
numerically \cite{benamar86}, imposing the calculated asymptotic
behaviour. The constant $B$ may be related to the nonlinear
eigenvalue $A$ by matching the outer WKB correction (\ref{transcorr})
with the transcendental contribution found in the inner domain
(\ref{Aeqn}): $A=2^{-\frac{3}{8}}\sigma^{-\frac{13}{28}} B.$ For the
tip slope, we get
\begin{equation}
\left. \frac{\mathrm{d}\eta_s}{\mathrm{d}\xi}\right\vert_{\xi=0}
=2^{\frac{11}{8}}\sigma^{-\frac{1}{28}}\,\mathrm{Im}(A) 
\exp\left\{\sigma^{-\frac{1}{2}}S(0)\right\} \>,\label{slope}
\end{equation}
a result that shows that with isotropic capillary length there is no
solution to the selection problem, as the right-hand side of
(\ref{slope}) is different from zero.

Let us now consider how the approach works when a potential flow is
imposed externally, with the nondimensional flow velocity ${\mathbf
  U}$ tending to $-P_f {\mathbf e}_y$ for $y\to\infty$.  We choose
this irrotational frictionless flow, because for this case the exact
solution of the problem without surface tension, i.e., the analog of
the Ivantsov solution, is known \cite{benamar88}.  Clearly, this is
just a toy model, as it cannot be used to study viscosity effects.
But it is a useful simple example in demonstrating our method,
extensible to more realistic flow patterns without difficulties of
principle.
Introducing 
the stream function $\psi$ via
$U_x=\partial_y \psi$, $U_y=-\partial_x \psi$, the bulk equation in
the liquid region, now {\em nonlinear}, reads \cite{benamar88}:
\begin{equation}
T_{\xi\xi}+T_{\eta\eta}=
\psi_\eta T_{\xi}-\psi_\xi T_{\eta}
+e^{-\frac{1}{2}P_f(\eta-1)^2}\psi_\xi.
\label{fieldliquid}
\end{equation}
Here, the stream function $\psi$ is determined by
$\psi_{\xi\xi}+\psi_{\eta\eta}=0$ with the boundary conditions
$\lim_{\eta\to\infty} \left(\psi-\psi_0\right) = 0$, where $\psi_0= P_f (\eta-1)
\xi$ is the Ivantsov-like solution for the stream function obtained
with $d_0=0$, and
\begin{equation}
\psi_\xi        
=-P_f(\xi h)^\prime
\label{psiboundary}
\end{equation}
at the interface.
The field equation in the solid region and the
interface equations  (\ref{balance}) remain unchanged.
As before, we use (\ref{bulks2}) 
in the solid. Analogically to (\ref{bulkl2}), we write 
\begin{equation}
(\partial_\xi+{i}\partial_\eta)\psi=0.\label{stream2}
\end{equation}
Equation (\ref{fieldliquid}), valid in the liquid region, does not
factorize (not even asymptotically), but we may use Zauderer's
asymptotic decomposition method \cite{zauderer78} to achieve a similar
reduction of order while keeping transcendentally small terms.  In
order not to overburden this first presentation, we linearize
(\ref{fieldliquid}) about $\psi_0$.  
Again, this is by no means a necessary step.  We have performed the
full analysis without this linearization, which leads to the same
equations (\ref{eq:Vsol}) and (\ref{eq:Wsol}) given below.
We first rewrite
equations (\ref{fieldliquid}) and (\ref{stream2}) together with the
boundary conditions (\ref{balance}) and
(\ref{psiboundary}), and using (\ref{bulks2}), as a first-order
system:
\begin{equation}
{\bf w}_\xi+A{\bf w}_\eta+
\epsilon \left\{B{\bf w}+C{\bf w}_\xi\right\}=0 \label{eqnsystem}
\end{equation}
with boundary condition
\begin{equation}\label{boundarysys}
\alpha{\bf w}+\beta{\bf w}_\xi={\bf b}\>,
\end{equation}  
where ${\bf w}=\left(T_\xi,T_\eta, \psi\right)^T$ and
\begin{eqnarray*}
{\bf b}=\left(\begin{array}{c}h^\prime-\frac{\sigma}{2}\kappa^\prime\\
i\frac{\sigma}{2}\kappa^\prime-(\xi h)^\prime\\-P_f(\xi h)^\prime
\end{array}\right), 
&& A=\left(\begin{array}{ccc}0&1&0\\-1&0&0\\0&0&i\end{array}\right)\>.
\end{eqnarray*}
Using the Kronecker symbol $\delta_{ij}$, the remaining matrices may
be written $B_{ij} =  \delta_{i1} \left[-P_f\xi \delta_{j1} + P_f(\eta-1)
  \delta_{j2} \right]$, $C_{ij} =
-e^{-\frac{1}{2}P_f(\eta-1)^2} \delta_{i1} \delta_{j3}$, $\alpha_{ij}
= \delta_{ij} (1-\delta_{i3})$, $\beta_{ij} = \delta_{ij}-
\alpha_{ij}$.
$\epsilon$ has been inserted into (\ref{eqnsystem}) as a bookkeeping
variable, to keep track of the order of Zauderer's decomposition
scheme. The formal procedure would be to write $\xi=-i+\epsilon X$,
$\eta=1+\epsilon Y$ and express the equations in terms of $X$ and $Y$.
In this presentation, we already decomposed the equations partially by
writing Eq.~(\ref{stream2}).
The parameter $\epsilon$ is assumed small as $\sigma\ll 1$. It is
possible to determine the scaling exponent for $\epsilon$ in a more
formal approach.  $A$ has the following eigenvectors:
\begin{equation}
{\bf r}_{11}=\left(\begin{array}{c}-i\\1\\0\end{array}\right),\quad
{\bf r}_{12}=\left(\begin{array}{c}0\\0\\1\end{array}\right),\quad
{\bf r}_{2}=\left(\begin{array}{c}i\\1\\0\end{array}\right),
\end{equation} 
with eigenvalues $i$, $i$, $-i$, respectively.
We look for a solution of the form
\begin{equation}\label{ansatz}
{\bf w}=\Vau{\bf r}_{11}+\psi{\bf r}_{12}+ \epsilon \We{\bf r}_{2}\>,
\end{equation}
assuming $ \epsilon\We$ to be much smaller than $\Vau$ and $\psi$.
Expanding the resulting system of equations to first order in
$\epsilon$, we obtain a set of decoupled first order equations.
Setting $\epsilon=1$ and introducing characteristic coordinates
$s=-i(\eta-1)$, $\tau=\xi+i(\eta-1)$, we find
$\psi=\psi(s,\tau)=-P_f[\tau h(\tau)]$, and using the boundary
conditions at the interface, we arrive at the two equations
\begin{eqnarray}
\Vau&=&\Vau(s,\tau)=\frac{i}{2}e^{\frac{P_f}{2}(s^2+\tau s)}
\Biggl\{\left[(1+i\tau)h(\tau)\right]^\prime \nonumber\\
&&\mbox{} -2\left[\tau h(\tau)\right]^\prime
\frac{1-e^{-\frac{P_f}{2}\tau s}}{\tau}\Biggr\} \>,\label{eq:Vsol}
\end{eqnarray}
\begin{eqnarray}
&&\We\mbox{}=\mbox{}\We(\bar{s},\bar{\tau})
=-\frac{P_f}{2}\bar{\tau}\int\limits_0^{\bar{s}}\mathrm{d}\bar{s}^\prime\,
\Vau(-\bar{s}^\prime,2\bar{s}^\prime+\bar{\tau})\nonumber\\
&&\mbox{}+i\frac{P_f}{2}\int\limits_0^{\bar{s}}\mathrm{d}\bar{s}^\prime\,
\left[h(2\bar{s}^\prime+\bar{\tau})+(2\bar{s}^\prime+\bar{\tau})
h^\prime(2\bar{s}^\prime+\bar{\tau})\right]
e^{\frac{P_f}{2}{\bar{s}^{\prime 2}}}\nonumber\\
&&\mbox{}+\frac{i}{2}\left\{\sigma\kappa^\prime(\bar{\tau})
-\left[(1-i\bar{\tau})h(\bar{\tau})\right]^\prime\right\},
\label{eq:Wsol}
\end{eqnarray}
where $\bar{s}=i(\eta-1), \bar{\tau}=\xi-i(\eta-1).$ 
Next, we require $\We$ to vanish for $\bar{s}\rightarrow i\infty$
while keeping $\bar{\tau}$ fixed.  $\bar{\tau}$ may be interpreted as
the continuation of the variable $\xi$ into the lower half of the
complex plane. Thus we write $\xi$ instead of $\bar{\tau}$, and after
some manipulations Eq.~(\ref{eq:Wsol}) yields
\begin{eqnarray}
&&(1-i\xi)h
+\frac{P_f}{4}e^{\frac{1}{8}P_f\xi^2}
\int\limits^\xi\mathrm{d}\xi^\prime\,e^{-\frac{1}{8}P_f\xi^{\prime 2}}
\Biggl[(1-i\xi^\prime)\xi^\prime h(\xi^\prime)\nonumber\\
&&\mbox{}+\frac{P_f}{2}\int\limits_{\xi^\prime}^{i\infty}\mathrm{d}
\xi^{\prime\prime}\,(\xi^{\prime\prime}
-\xi^{\prime})\xi^{\prime\prime}h(\xi^{\prime\prime})
e^{\frac{P_f}{8}(\xi^{\prime\prime}-\xi^{\prime})^2}\Biggr]
=\sigma\kappa.\label{LE2-} 
\end{eqnarray}
For $P_f=0$, Eq.~(\ref{LE2-}) reduces to (\ref{basic-}).
The calculation of the transcendental mismatch in the presence of a
potential flow parallels the procedure for the flowless case.  We first
calculate the WKB solution of (\ref{LE2-}):
\begin{equation}
h=B_1 e^{\frac{P_f}{16}}(1-{i}\xi)^{-\frac{5}{8}}
(1+{i}\xi)^{-\frac{3}{8}}
\exp\left\{\sigma^{-\frac{1}{2}}S(\xi)+\frac{P_f}{16}\xi^2\right\}
\label{transcorr1}
\end{equation}
with $S(\xi)$ given in Eq.~(\ref{phaseWKB}).  To obtain the inner
equation, we follow Ben Amar \cite{benamar90} 
in her treatment of the flowless finite growth Péclet number case 
and use the same scaling as in the case without convection, i.e.,
Eq.~(\ref{rescaling}) with $\alpha=\frac{2}{7}$. Equation (\ref{LE2-})
simplifies to
\begin{equation}\label{innerwithflow}
 t\phi +P_1\int\limits^t\mathrm{d}t^\prime\,t^\prime \phi(t^\prime) = K \>,
\end{equation}
with $K$ defined in Eq.~(\ref{Kdef}) and
$P_1=\sigma^\frac{2}{7}\frac{P_f}{4}$. 
To leading
asymptotic order, 
\begin{equation}
P_1\int\limits^t\mathrm{d}t^\prime\,t^\prime \phi(t^\prime)
\sim 2^{-\frac{3}{2}}t^{-\frac{3}{2}}, \quad (t \gg 1). \label{hflowlead}
\end{equation}
Linearizing equation (\ref{innerwithflow}) about (\ref{hflowlead})
and performing a local asymptotic analysis near $t=\infty$ we obtain
\begin{equation}
g=t^{-\frac{5}{8}}\exp\left\{\frac{P_1}{2}t\right\}
\exp\left\{-\frac{4}{7}2^{\frac{1}{4}}t^{\frac{7}{4}}\right\}\>,
\end{equation}
leading to
\begin{equation}
\phi = 2^{-\frac{3}{2}}t^{-\frac{5}{2}}+A_1(P_1)\,  
g(t)\,\,\,\,\,\,\,(t\rightarrow\infty) \>,\label{A1}
\end{equation}
where  $A_1(P_1)$ is a nonlinear eigenvalue and a function of $P_1$. 
Matching (\ref{transcorr1}) and (\ref{A1}), we obtain the tip slope
\begin{equation}
\left. \frac{\mathrm{d}\eta_s}{\mathrm{d}\xi}\right\vert_{\xi=0}
=2^{\frac{11}{8}}\sigma^{-\frac{1}{28}}\mathrm{Im}\left\{A_1\right\} 
\exp\left\{\frac{P_f}{16}\right\}\exp\left\{\sigma^{-\frac{1}{2}}S(0)\right\}.
\label{slopePf}
\end{equation}
Obviously, the mismatch will remain nonzero except possibly for
isolated values of $P_1$, a case that may be excluded by numerical
evaluation of $A_1(P_1)$.  Hence, there is no admissible solution.

To include surface tension anisotropy, we have to replace $\kappa$ in
 (\ref{LE2-}) with $R\kappa$, where
\begin{equation}
R=1-\beta\cos(4\theta)
=1-\beta\cos\left[4\arctan\left|\frac{\eta_s\eta_s^\prime-\xi}{\eta_s
+\xi\eta_s^\prime}\right|\right].
\end{equation}
Studying the vicinity of $\xi= -i$, we find
\begin{equation}\label{innerfinal}
t\phi+P_1\int\limits^t\mathrm{d}t^\prime\,t^\prime \phi(t^\prime)=K+bHK \>,
\end{equation}
where 
$ H=-2(\dot{\phi}-1)^2/\left[(\dot{\phi}+1)^2(\phi+t)^2\right] $
and $b=\beta\sigma^{-\frac{4}{7}}$. 
Demanding $\mathrm{Im}(A_1)$ to vanish, equation (\ref{innerfinal})
constitutes an eigenvalue problem for the parameter $b$, to be solved
numerically. Denoting the lowest eigenvalue by $b=b_0(P_1)$ the
selection criterion reads
\begin{equation}\label{sigmastar}
\sigma=f(P_1)\beta^{7/4},\;\;\;\;f(P_1)=\left[b_0(P_1)\right]^{-7/4}.
\end{equation}
We will not elaborate on the details of this solution, as the purpose
of this article is only a demonstration of the method.  Further results for
the particular physical system will be discussed in a forthcoming paper.


To summarize, we have introduced a method that combines 
matched asymptotics in the complex plane with the asymptotic
decomposition of partial differential equations.  This allows one to
compute exponentially small terms beyond all orders (of asymptotic
expansions) for partial differential equations on free boundaries, as
we have shown for dendritic growth in a forced potential flow.  In
comparison with the BP approach~\cite{bouissou89}, ours has several
advantages, the most important being that it paves the way for a
rigorous nonlinear asymptotic analysis, which in some cases
\cite{tanveer90} seems to be the only one that gives even
qualitatively correct answers. We are not aware of any other method
allowing this type of analysis with nonlinear field equations.
Many problems to which the Kruskal-Segur method has been applied so
far, including viscous fingering \cite{combescot86}, the $\psi^4$
breather of particle physics \cite{segur87}, or capillary water waves
\cite{pomeau88}, are actually free boundary problems. But for the
method to be applicable, they first had to be recast as an
ordinary differential equation or at least a single
differential-integral equation. This is the reason, why only the 
simplest, in some cases even the most unrealistic, physical situations,
have been studied by this method.  Without the restriction, we
anticipate a much wider class of problems to be tractable.  Hence we
expect our approach to open a new line of research into a plethora of
hitherto untractable selection or solvability problems not only in
crystal growth and similar problems of pattern forming
interface dynamics (such as viscous fingering), but in a vast number
of situations reducible to free boundary problems.

\newpage

\begin{thebibliography}{99}
\bibitem{expla} This linearization allows us to keep formulas less
  cumbersome. It should be avoided in a theory that is to become
  quantitatively accurate in the limit $d_0\to 0$.  A presentation of
  the slightly more involved full calculation might obscure the
  essence of the approach and is postponed to a more technical
  extended paper. The result (\ref{localeqn2}) remains exact in the
  full theory.
\end{thebibliography}
\end{document}